\documentclass[12pt]{article}
\usepackage{graphicx}
\usepackage{mcite}
\usepackage{units}

\newcommand\pubdate{\today}
                                                                                                                         
\newcommand\pubnumber{}

\def\Title#1{\begin{center} {\Large #1 } \end{center}}
\def\Author#1{\begin{center}{ \sc #1} \end{center}}
\def\Address#1{\begin{center}{ \it #1} \end{center}}

\newcommand\pubblock{\rightline{\begin{tabular}{l} \pubnumber\\
         \pubdate  \end{tabular}}}
\newenvironment{Abstract}{\begin{center}{\bf Abstract}\end{center} \bigskip \begin{quotation}  }{\end{quotation}}
\newenvironment{Presented}{\begin{quotation} \begin{center} 
             PRESENTED AT\end{center}\bigskip 
      \begin{center}\begin{large}}{\end{large}\end{center} \end{quotation}}

\def\beq{\begin{equation}}
\def\eeq#1{\label{#1}\end{equation}}
\def\eeqn{\end{equation}}

\def\beqa{\begin{eqnarray}}
\def\eeqa#1{\label{#1}\end{eqnarray}}
\def\eeqan{\end{eqnarray}}

\let\bar=\overbar

\def\Dslash{\not{\hbox{\kern-4pt $D$}}}
\def\dslash{\not{\hbox{\kern-2pt $\del$}}}

\def\msb{{\bar{\ssstyle M \kern -1pt S}}}

\begin{document}
\begin{titlepage}
\pubblock

\vfill


\Title{The current status of neutrino mixing}
\vfill
\Author{Justin J. Evans}
\Address{Department of Physics and Astronomy, University College London, Gower Street, London, WC1E 6BT, UK}
\vfill


\begin{Abstract}
A brief review of the experimental status of neutrino mixing.
The model of neutrino oscillations has now been established with high confidence, with many of the model parameters measured to an accuracy of a few per cent.
However, some parameters still remain unknown, notably the mixing angle $\theta_{13}$ and the amount of CP violation.
Recently, new questions have come to light, highlighting possibilities to search for new physics in the neutrino sector.
\end{Abstract}

\vfill

\begin{Presented}
The Ninth International Conference on\\
Flavor Physics and CP Violation\\
(FPCP 2011)\\
Maale Hachamisha, Israel,  May 23--27, 2011
\end{Presented}
\vfill

\end{titlepage}
\def\thefootnote{\fnsymbol{footnote}}
\setcounter{footnote}{0}
%


\section{Introduction}

In the 1960s, Ray Davis set up a 390,000-litre tank of dry
cleaning fluid in the Homestake mine in South Dakota, with
the aim of observing electron neutrinos produced by the Sun. He saw
significantly fewer than were predicted by solar
models~\cite{ref:HomestakeFirstResults}. This puzzle remained for two
decades until, in the late 1990s, the Super-Kamiokande
experiment in Japan showed conclusively that neutrinos disappeared as
they traveled~\cite{ref:SuperKFirstZenithAngle}. Super-Kamiokande looked at
muon neutrinos produced in the Earth's atmosphere. Neutrinos coming
from above, which had only traveled the thickness of the atmosphere,
showed the expected rate. Neutrinos from below, which had traveled the
full diameter of the Earth, were significantly depleted.

In the early years of the 21st century, the SNO collaboration looked
at electron neutrinos produced by the Sun~\cite{ref:SNODayNight}. They saw the same
depletion that Ray Davis had observed. However, the SNO experiment was
also able to measure the neutral current interaction rate, which
is independent of neutrino flavour. No depletion was seen in the neutral current
event rate, confirming that the missing electron neutrinos were
transforming into another flavour of neutrino.

\section{Neutrino oscillations}

This process of neutrino flavour change is what we now know to be
neutrino oscillation. The three states of neutrino mass do not
correspond to the states of neutrino flavour. A neutrino is created,
for example in the Sun or a nuclear reactor, in a state of definite
flavour. When it propagates, the flavour state splits into its
constituent mass states. The relative phases of these mass states
change during propagation, so that upon detection the neutrino is no
longer in a state of definite flavour, hence the detected neutrino can
have a different flavour from the neutrino produced at the source.

\begin{figure}[htb]
\centering
\includegraphics[width=0.6\textwidth]{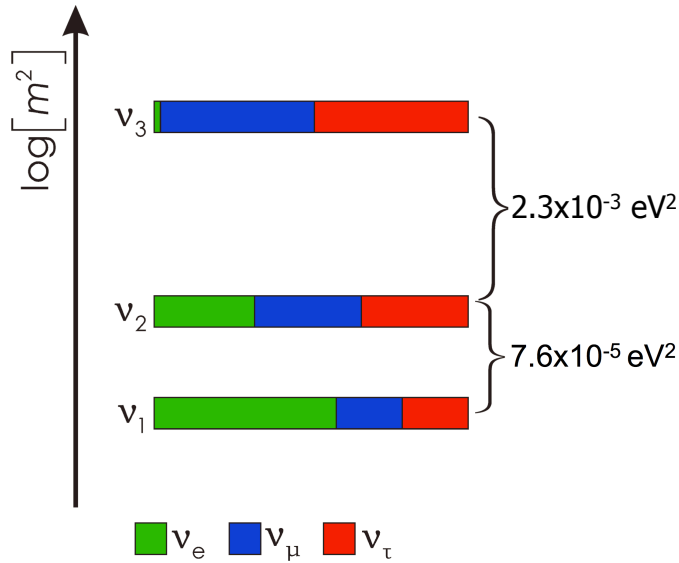}
\caption{A pictorial representation of the mixing between neutrino mass
  and flavour states. The horizontal bars represent the three neutrino
  mass states. The coloured bars represent the contribution of each
  neutrino flavour state to the mass state.}
\label{fig:MixingCartoon}
\end{figure}

The rate of the oscillations between flavours is governed by the
differences in the squared masses between the three neutrino mass
states: $\Delta m^{2}_{21}$ and $\Delta m^{2}_{32}$. The magnitude of
the oscillations is governed by the degree of mixing between the mass
and flavour states. This mixing is governed by the three-dimensional
PMNS matrix, which is parameterized by three mixing angles
$\theta_{12}$, $\theta_{13}$ and $\theta_{23}$, and by a phase
$\delta_{CP}$ which governs the amount of CP violation in the neutrino
sector. The mixing is shown schematically in Fig.~\ref{fig:MixingCartoon}.

\section{The solar sector}

The smaller of the two mass splittings, $\Delta m^{2}_{21}$, is often
referred to as the solar mass splitting. It drives oscillations of
neutrinos with a ratio of distance traveled to energy of the order of
$\unit[10^{5}]{km / GeV}$, so is well suited to be studied with solar
electron neutrinos and electron antineutrinos from nuclear
reactors. The level of disappearance of these electron neutrinos is
governed by the mixing angle $\theta_{12}$.

Amongst the many experiments to have studied this sector are the SNO
experiment~\cite{ref:SNONIM}, which is situated in the Sudbury nickel
mine in Canada, and has ceased taking data in its current form. It
consisted of a \unit[1]{kt} tank of heavy water, viewed by
photomultiplier tubes, observing the Cerenkov radiation of the
products of neutrino interactions.

The KamLAND experiment in Japan~\cite{ref:KamLANDdetector} is a \unit[1]{kt} tank of
liquid scintillator, again viewed by photomultiplier tubes. KamLAND
looks at the electron antineutrinos produced by the many nuclear
reactors which surround it, typically at a distance of \unit[180]{km}.

\begin{figure}[htb]
\centering
\includegraphics[width=0.6\textwidth]{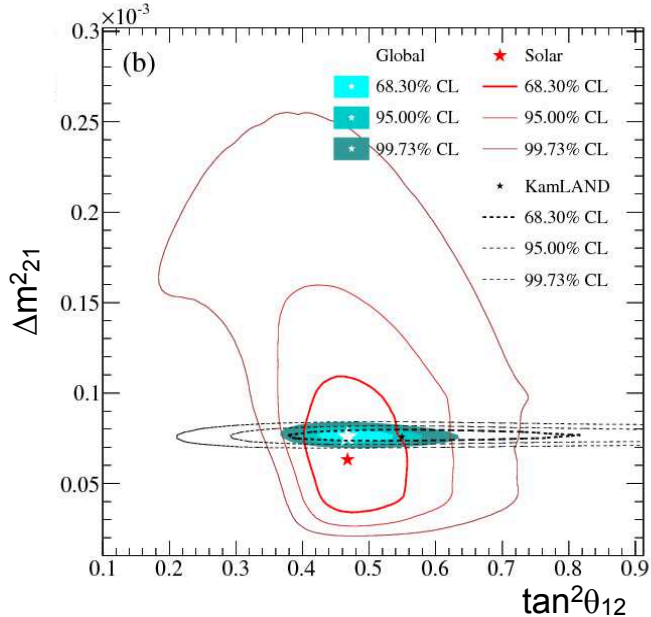}
\caption{A summary of all constraints on the solar oscillation
  parameters from~\protect\cite{ref:GlobalSolarLimits}. These limits are obtained from a fit using a
  three-neutrino model.}
\label{fig:SolarKnowledge}
\end{figure}

The current state of knowledge of the solar oscillation parameters is
summarized in Fig.~\ref{fig:SolarKnowledge}.
The constraints labeled `solar' include all relevant data: Borexino~\cite{ref:Borexino},
the gallium~\cite{ref:Gallium} and chlorine~\cite{ref:Chlorine} experiments, Super-Kamiokande~\cite{ref:SuperKSolar1,ref:SuperKSolar2} and the most
recent analysis from the SNO collaboration in which significant work
was done to lower the energy threshold of the
experiment~\cite{ref:GlobalSolarLimits}.
 The solar experiments provide
the strongest constraints on the mixing angle $\theta_{12}$; the
KamLAND reactor data provides the strongest constraints on the mass
splitting $\Delta m^{2}_{21}$~\cite{ref:KamLANDbestLimits}. The fact
that the solar oscillations are strongly affected by the dense matter
in the Sun's interior further allows the sign of the mass splitting to
be resolved and the value of $\theta_{12}$ to be uniquely
determined. (Oscillations in vacuum provide no constraints on the sign
of the mass splitting, and leave the quadrant of the mixing angle
ambiguous).

The combination of the global data yields $\Delta m^{2}_{21} = \unit[7.59^{+0.20}_{-0.21}\times 10^{-5}]{eV^{2}}$ and $\theta_{12} = (34.06^{+1.16}_{-0.84})^{\circ}$~\cite{ref:GlobalSolarLimits}.

\section{The atmospheric sector}

The larger mass splitting, $\Delta m^{2}_{32}$, is often known as the atmospheric mass splitting.
It is well suited to being studied using muon neutrinos with a ratio of distance traveled to energy of $\unit[10^{3}]{km / GeV}$.
The amount of muon neutrino disappearance over this distance is governed by the mixing angle $\theta_{23}$.

The MINOS experiment~\cite{ref:MINOSNIM} uses a beam of muon neutrinos, produced at the Fermilab accelerator complex in Chicago.
Two detectors measure the energy spectrum of the muon neutrinos: one at Fermilab measures the spectrum before oscillations have occurred.
A second detector, as similar as possible to the first, is located \unit[735]{km} from the source, at the Soudan underground laboratory in Minnesota.
This measures the energy spectrum after the oscillations have time to manifest.
The comparison of measurements from the two spectra is very powerful for the mitigation of systematic uncertainties, since many sources of uncertainty, such as mismodeling of the neutrino flux or cross sections, affect both detectors in the same way, so cancel in the detector-to-detector comparison. MINOS measures the oscillation parameters as $\Delta m^{2}_{32} = \unit[2.32^{+0.12}_{-0.08}\times 10^{-3}]{eV^{2}}$ and $\sin^{2}(2\theta_{23}) > 0.90$ (90\% confidence limit), making the world's best measurement of the mass splitting~\cite{ref:MINOSNeutrinoLimits}.

The Super-Kamiokande experiment~\cite{ref:SuperKNIM} is a \unit[50]{kt} water Cerenkov detector, looking for the disappearance of muon neutrinos produced in the Earth's atmosphere.
Super-Kamiokande makes the world's best measurement of the mixing angle, measuring $\Delta m^{2}_{32} = \unit[2.11^{+0.11}_{-0.19}\times 10^{-3}]{eV^{2}}$ and $\sin^{2}(2\theta_{23}) > 0.96$ (90\% confidence limit)~\cite{ref:SuperKNeutrino2010}.

\begin{figure}[htb]
\centering
\includegraphics[width=0.6\textwidth]{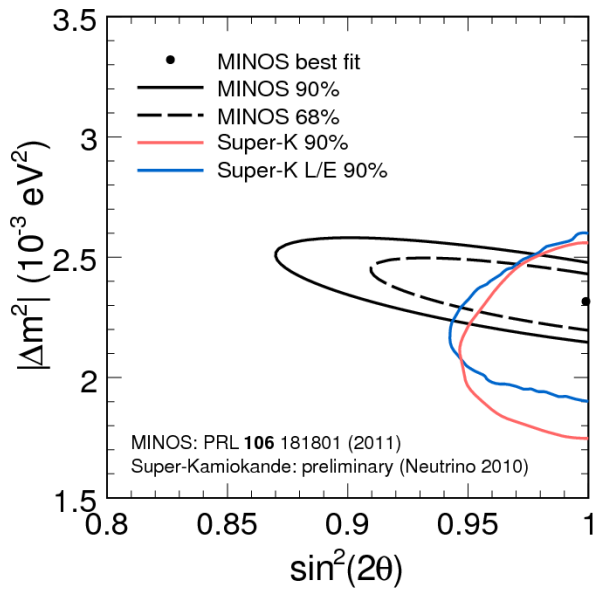}
\caption{Limits on the atmospheric oscillation parameters from the
  MINOS accelerator neutrino experiment and the Super-Kamiokande
  atmospheric neutrino experiment.}
\label{fig:AtmosphericParameters}
\end{figure}

The MINOS and Super-Kamiokande measurements are shown in Fig.~\ref{fig:AtmosphericParameters}.

\section{Antineutrinos}

The rate of muon neutrino disappearance in the atmospheric sector should be identical between neutrinos and antineutrinos.
Any difference between the two would indicate new physics.

\begin{figure}[htb]
\centering
\includegraphics[width=0.6\textwidth]{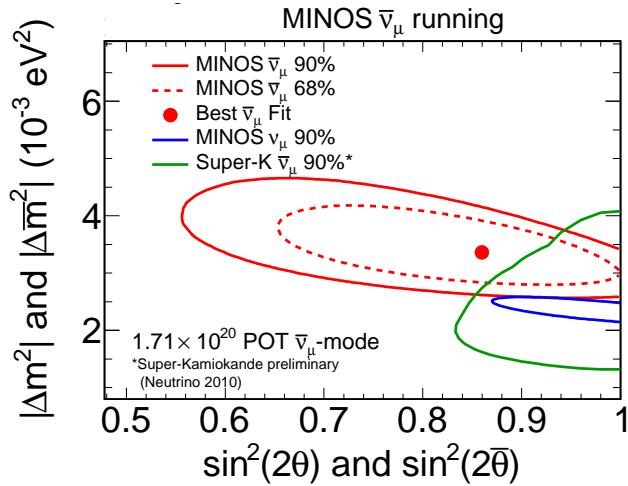}
\caption{Limits on the antineutrino oscillation parameters in the
  atmospheric regime from the MINOS~\protect\cite{ref:MINOSRHCPRL} and Super-Kamiokande
  experiments~\protect\cite{ref:SuperKNeutrino2010}. For comparison, the MINOS neutrino limit~\protect\cite{ref:MINOSNeutrinoLimits} is shown in
  blue.}
\label{fig:AntineutrinoParameters}
\end{figure}

The MINOS experiment has the ability to distinguish muon neutrinos and
antineutrinos on an event-by-event basis since the detectors are
magnetized: the sign of the charge of the muon produced in charged
current muon neutrino interactions can be measured. MINOS has taken
data with a dedicated muon antineutrino beam; the results of that
measurement are shown in Fig.~\ref{fig:AntineutrinoParameters}.
MINOS measures $\Delta\overline{m}^{2}_{32} = \unit[3.36^{+0.46}_{-0.40}\mathrm{(stat.)}\pm0.06\mathrm{(syst.)}\times 10^{-3}]{eV^{2}}$ and $\sin^{2}(2\overline{\theta}_{23}) = 0.86^{+0.11}_{-0.12}\mathrm{(stat.)}\pm0.01\mathrm{(syst.)}$~\cite{ref:MINOSRHCPRL}.
The MINOS neutrino and antineutrino measurements are consistent at the 2.0\% confidence level, assuming identical underlying oscillation parameters.

The Super-Kamiokande experiment cannot separate neutrino and antineutrino interactions on an event-by-event basis.
However, by knowing the relative contribution of neutrinos and antineutrinos to the atmospheric flux, Super-Kamiokande can statistically constrain the antineutrino oscillation parameters.
The Super-Kamiokande measurement is also shown in Fig.~\ref{fig:AntineutrinoParameters}, and is the best measurement of the antineutrino mixing angle~\cite{ref:SuperKNeutrino2010}.
The Super-Kamiokande measurement of the antineutrino mass splitting is consistent with both the MINOS neutrino and antineutrino measurements.

\section{Tau neutrino appearance}

The neutrino oscillation model states that the muon neutrinos disappearing through oscillations in the atmospheric sector are predominantly turning into tau neutrinos.
However, this tau neutrino appearance has never been observed.
The OPERA experiment aims to make this observation, using a beam of muon neutrinos produced at CERN, traveling over \unit[730]{km}.
The OPERA detector is a fine-grained emulsion detector.
The charged current interaction of a tau neutrino produces a tau lepton, which travels a few millimetres before decaying, leaving a characteristic kinked track visible in the emulsion.

\begin{figure}[htb]
\centering
\includegraphics[width=\textwidth]{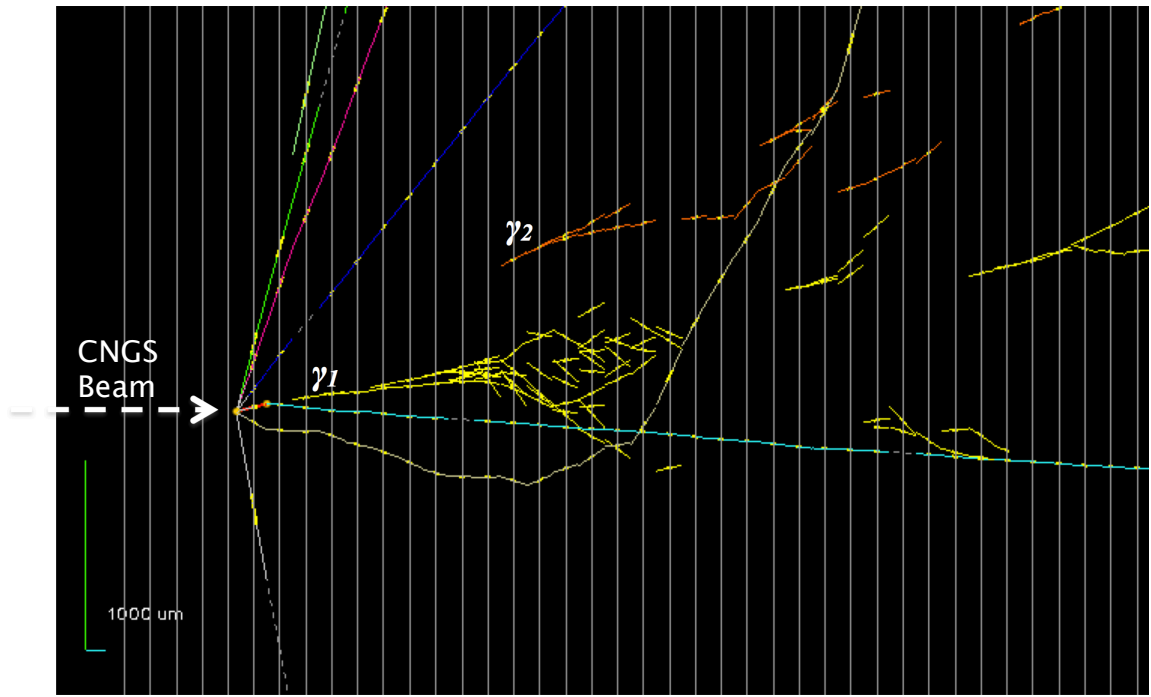}
\caption{The candidate $\nu_{\tau}$ interaction observed by the OPERA
  experiment. The short red line represents the candidate tau lepton
  track, which quickly decays to a negative pion (light blue line) and
  a neutral pion (which leaves no signature). The neutral pion decays
  to two photons, producing the two electromagnetic showers labeled
  $\gamma_{1}$ and $\gamma_{2}$.}
\label{fig:OperaEvent}
\end{figure}

In the summer of 2010, OPERA observed the event shown in Fig.~\ref{fig:OperaEvent}~\cite{ref:OperaEvent}.
This event is consistent with a charged current tau neutrino interaction: the tau lepton decaying into a positive pion and a neutral pion, the latter not visible in the detector.
The neutral pion then decays into two photons, leaving the two electromagnetic showers visible further downstream.
The probability that this event is not background is 2.01 standard deviations.

\section{The mixing angle $\theta_{13}$}

Although most muon neutrinos are converted to tau neutrinos by atmospheric-sector oscillations, a small fraction may be converting to electron neutrinos.
This fraction is governed by the mixing angle $\theta_{13}$ which has not yet been measured, but which is known to be small.
The importance of measuring this angle lies in the fact that CP violation in the neutrino sector can only be observed if $\theta_{13}$ is non-zero.

\begin{figure}[!htb]
\centering
\includegraphics[width=0.6\textwidth]{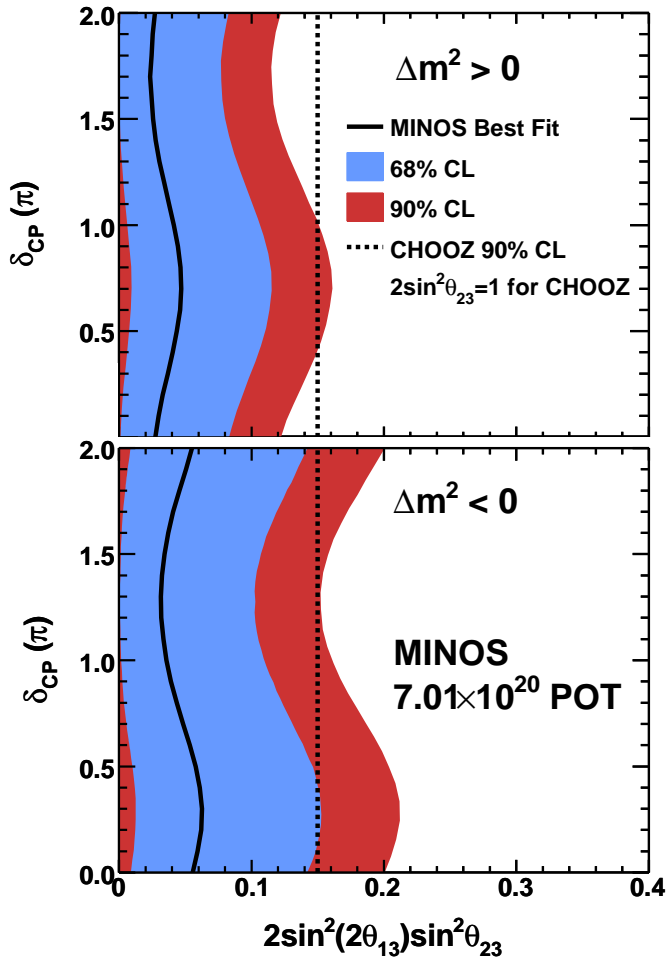}
\caption{Limits on the mixing angle $\theta_{13}$ from the MINOS experiment~\protect\cite{ref:MINOSNuE}. The previous best limit from the CHOOZ experiment~\protect\cite{ref:CHOOZ} is indicated by the vertical black line.}
\label{fig:ThetaOneThree}
\end{figure}

The MINOS experiment has set limits on $\theta_{13}$, which are summarized in figure~\ref{fig:ThetaOneThree}~\cite{ref:MINOSNuE}.
These limits are compared to the previous best measurement: that of the CHOOZ collaboration which searched for the disappearance of electron antineutrinos produced by a nuclear reactor~\cite{ref:CHOOZ}.
The MINOS limits depend on two unknowns: the sign of $\Delta m^{2}_{32}$ and the phase $\delta_{CP}$.
The limits are also dependent on the value of the mixing angle $\theta_{23}$.
Assuming $\delta_{CP} = 0$, $\sin^{2}(2\theta_{23}) = 1.0$ and $|\Delta m^{2}_{32}| = \unit[2.43\times10^{-3}]{eV^{2}}$, MINOS constrains $\sin^{2}(2\theta_{13}) < 0.12$ (90\% confidence limit) for a positive $\Delta m^{2}_{32}$ and $\sin^{2}(2\theta_{13}) < 0.20$ (90\% confidence limit) for a negative $\Delta m^{2}_{32}$.

\section{Sterile neutrinos}

In the 1990s, the LSND experiment saw evidence for oscillations driven by a mass splitting of the order of $\unit[1]{eV^{2}}$~\cite{ref:LSNDResults}.
This would require the existence of a fourth neutrino state.
Measurements of the decay width of the $Z$ boson show that only three light active neutrino flavours exist~\cite{ref:pdg}.
Any additional neutrinos therefore have to be sterile: not experiencing the weak interaction.

The MiniBooNE experiment~\cite{ref:MiniBooNENIM} was constructed to directly test the LSND observation.
MiniBooNE looked for electron neutrino appearance in a muon neutrino beam.
No evidence was seen for such appearance in the LSND signal region, apparently refuting the LSND claim~\cite{ref:MiniBooNENeutrinos}.
At energies below the signal region, an excess with a significance of three standard deviations was observed.
After careful checks for potential systematic sources for this excess, no candidates have been identified, so the excess remains unexplained.

\begin{figure}[htb]
\centering
\includegraphics[width=0.6\textwidth]{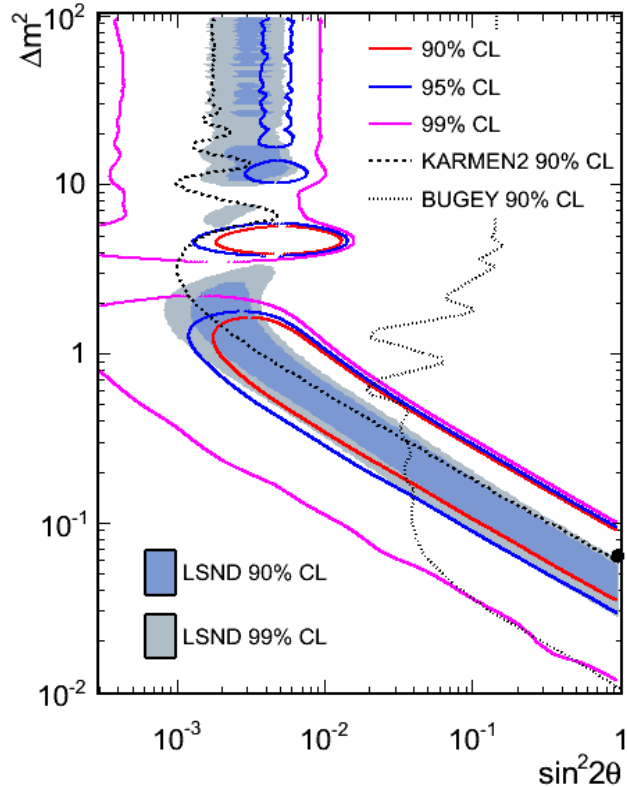}
\caption{The allowed region for antineutrino oscillations driven by a mass splitting of the order of $\unit[1]{eV^{2}}$.}
\label{fig:LSND}
\end{figure}

However, the LSND collaboration performed their measurements using antineutrinos.
The MiniBooNE collaboration therefore repeated their experiment with antineutrinos, observing an excess of electron neutrinos in the LSND signal region, the excess being consistent with no appearance at the 3\% confidence level~\cite{ref:MiniBooNEAntineutrinos}.
The allowed region for the oscillation parameters implied by the LSND and MiniBooNE electron antineutrino appearance searches is shown in Fig.~\ref{fig:LSND}; with a full fit, MiniBooNE excludes the null hypothesis at the 99.4\% confidence level.

\begin{figure}[htb]
\centering
\includegraphics[width=0.6\textwidth]{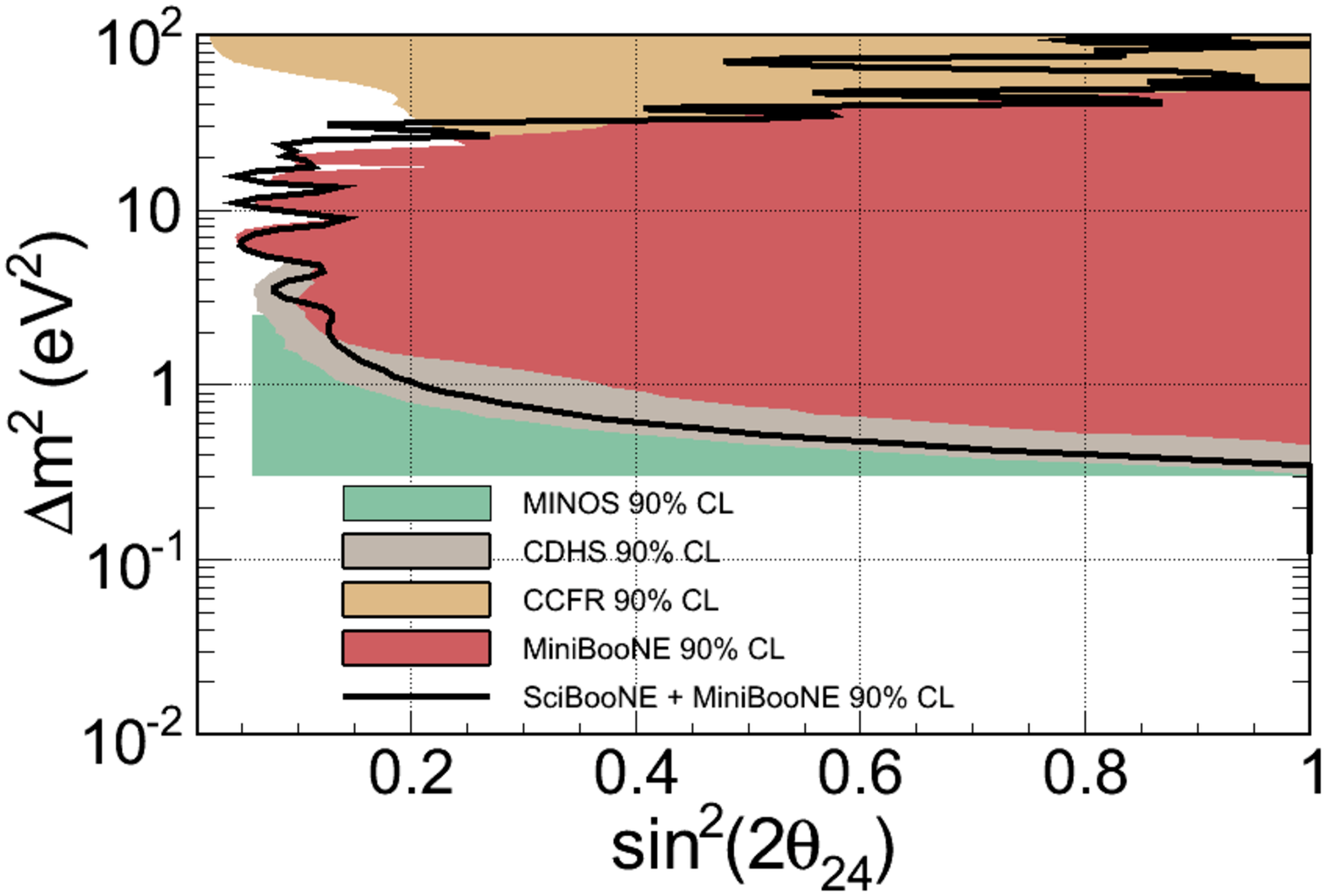}
\caption{Limits on oscillations driven by a mass splitting of the order of $\unit[1]{eV^{2}}$ as measured using the $\nu_{\mu}$ disappearance channel. The MINOS limit uses data from ~\protect\cite{ref:MINOSNCPRL}; the other limits are taken from~\protect\cite{ref:MiniBooNEDisappearance}.}
\label{fig:MINOSSteriles}
\end{figure}

Oscillations to a sterile neutrino would also cause a disappearance of muon neutrinos over the same baseline to energy ratio.
The limits arising from this channel (which is complimentary to the electron neutrino appearance channel, since it probes a different combination of mixing angles) are summarized in Fig.~\ref{fig:MINOSSteriles}.
No evidence for sterile neutrinos is seen in this channel.

A recent re-evaluation of the electron antineutrino flux from nuclear reactors suggests an increase of the expected flux with respect to previous estimations~\cite{ref:ReactorAnomaly}.
If correct, this would mean the previous short baseline reactor experiments have, on average, been observing a 5.7\% deficit.
This deficit can be interpreted to be caused by oscillations to a sterile neutrino, again driven by a mass splitting of the order of $\unit[1]{eV^{2}}$.

\section{Summary}

Neutrino mixing is now a mature field, the model of neutrino oscillation established with high confidence.
Many of the parameters of the oscillation framework have been measured to precisions of a few per cent.
However, many unknowns remain: the mixing angle $\theta_{13}$, the sign of the atmospheric mass splitting, and the amount of CP violation in the neutrino sector.
In recent years, new questions have come to prominence which have the potential to probe new physics: the equivalence of the neutrino and antineutrino oscillation rates, and the possible existence of sterile neutrinos.
An exciting future therefore awaits the field, as an array of current and future experiments continue to probe these unanswered questions over the coming decades.

\bibliography{References}

\begin{thebibliography}{10}

\bibitem{ref:HomestakeFirstResults}
R.~Davis, D.~S. Harmer, and K.~C. Hoffman,
\newblock Phys. Rev. Lett. {\bf 20}, 1205 (1968).

\bibitem{ref:SuperKFirstZenithAngle}
Y.~Fukuda et~al.,
\newblock Phys. Rev. Lett. {\bf 81}, 1562 (1998).

\bibitem{ref:SNODayNight}
B.~Aharmim et~al.,
\newblock Phys. Rev. {\bf C72}, 055502 (2005).

\bibitem{ref:SNONIM}
J.~Boger et~al.,
\newblock Nucl.\ Instrum.\ Meth. {\bf A449}, 172 (2000).

\bibitem{ref:KamLANDdetector}
S.~Abe et~al.,
\newblock Phys.\ Rev. {\bf C81}, 025807 (2010).

\bibitem{ref:Borexino}
C.~Arpesella et~al.,
\newblock Phys.\ Rev.\ Lett. {\bf 101}, 091302 (2008).

\bibitem{ref:Gallium}
J.~N. Abdurashitov et~al.,
\newblock Phys.\ Rev. {\bf C80}, 015807 (2009).

\bibitem{ref:Chlorine}
B.~T. Cleveland et~al.,
\newblock Astrophys. J. {\bf 496}, 505 (1998).

\bibitem{ref:SuperKSolar1}
J.~Hosaka et~al.,
\newblock Phys.\ Rev.\ {\bf D73}, 112001 (2006).

\bibitem{ref:SuperKSolar2}
J.~P. Cravens et~al.,
\newblock Phys.\ Rev. {\bf D78}, 032002 (2008).

\bibitem{ref:GlobalSolarLimits}
B.~Aharmim et~al.,
\newblock Phys.\ Rev. {\bf C81}, 055504 (2010).

\bibitem{ref:KamLANDbestLimits}
S.~Abe et~al.,
\newblock Phys.\ Rev.\ Lett {\bf 100}, 221803 (2008).

\bibitem{ref:MINOSNIM}
D.~G. Michael et~al.,
\newblock Nucl.\ Instrum.\ Meth. {\bf A569}, 190 (2008).

\bibitem{ref:MINOSNeutrinoLimits}
P.~Adamson et~al.,
\newblock Phys.\ Rev\ Lett. {\bf 106}, 181801 (2011).

\bibitem{ref:SuperKNIM}
S.~Fukuda et~al.,
\newblock Nucl.\ Instrum.\ Meth. {\bf A501}, 418 (2003).

\bibitem{ref:SuperKNeutrino2010}
Y.~Takeuchi,
\newblock (2010),
\newblock Contributed to XXIV international conference on neutrino physics and
  astrophysics (Neutrino 2010), Athens, Greece, 14--19 June 2010.

\bibitem{ref:MINOSRHCPRL}
P.~Adamson et~al.,
\newblock (2011),
\newblock Accepted for publication by Phys.\ Rev.\ Lett.

\bibitem{ref:OperaEvent}
N.~Agafonova et~al.,
\newblock Phys.\ Lett. {\bf B691}, 138 (2010).

\bibitem{ref:MINOSNuE}
P.~Adamson et~al.,
\newblock Phys. Rev. {\bf D82}, 051102 (2010).

\bibitem{ref:CHOOZ}
M.~Apollonio et~al.,
\newblock Eur. Phys. J. {\bf C27}, 331 (2003).

\bibitem{ref:LSNDResults}
A.~Aguilar et~al.,
\newblock Phys. Rev. {\bf D64}, 112007 (2001).

\bibitem{ref:pdg}
K.~Nakamura et~al.,
\newblock J. Phys. {\bf G37}, 075021 (2010).

\bibitem{ref:MiniBooNENIM}
A.~A. Aguilar-Arevalo et~al.,
\newblock Nucl.\ Instrum.\ Meth. {\bf A599}, 28 (2009).

\bibitem{ref:MiniBooNENeutrinos}
A.~A. Aguilar-Arevalo et~al.,
\newblock Phys.\ Rev.\ Lett. {\bf 102}, 101802 (2009).

\bibitem{ref:MiniBooNEAntineutrinos}
A.~A. Aguilar-Arevalo,
\newblock Phys.\ Rev.\ Lett. {\bf 105}, 181801 (2010).

\bibitem{ref:MINOSNCPRL}
P.~Adamson et~al.,
\newblock (2011),
\newblock Accepted for publication by Phys.\ Rev.\ Lett.

\bibitem{ref:MiniBooNEDisappearance}
A.~A. Aguilar-Arevalo et~al.,
\newblock Phys.\ Rev.\ Lett. {\bf 103}, 061802 (2009).

\bibitem{ref:ReactorAnomaly}
G.~Mention et~al.,
\newblock Phys.\ Rev. {\bf D83}, 073006 (2011).

\end{thebibliography}





\end{document}